# An H I study of the collisional ring galaxy NGC 922

Ahmed Elagali,[1][*] O. Ivy Wong,[1,2] Se-Heon Oh,[1,3] Lister Staveley-Smith,[1,2] Bärbel S. Koribalski,[4] Kenji Bekki[1] and Martin Zwaan[5]

[1]*International Centre for Radio Astronomy Research (ICRAR), M468, The University of Western Australia, 35 Stirling Highway, Crawley, WA 6009, Australia*
[2]*Australian Research Council Centre of Excellence for All-sky Astrophysics (CAASTRO), Australia*
[3]*Korea Astronomy and Space Science Institute (KASI), Daejeon 305-348, Korea*
[4]*CSIRO Astronomy and Space Science, Australia Telescope National Facility, PO Box 76, Epping, NSW 1710, Australia*
[5]*European Southern Observatory, Karl-Schwarzschild-Strasse 2, D-85748 Garching bei München, Germany*

**ABSTRACT**

We present new atomic hydrogen (H I) observations of the collisional ring galaxy NGC 922 obtained using the Australia Telescope Compact Array. Our observations reveal for the first time the vast extent of the H I disc of this galaxy. The H I morphology and kinematics of NGC 922 show that this galaxy is not the product of a simple drop-through interaction, but has a more complex interaction history. The integrated H I flux density of NGC 922 from our observations is 24.7 Jy km s$^{-1}$, which is within the error of the flux value obtained using the 64-m Parkes radio telescope. This flux density translates to a total H I mass of $1.1 \times 10^{10}$ M$_\odot$ and corresponds to an H I to total mass fraction ($M_{\mathrm{H I}}/M_{\mathrm{tot}}$) of approximately 0.11. The gaseous structures of NGC 922 are more extended to the north and include an H I tail that has a projected physical length of 8 kpc. Gas warps are also evident in the velocity field of NGC 922 and are more prominent on the approaching and the western side of the disc. In comparison with a large sample of star-forming galaxies in the local Universe, NGC 922 possesses a high gas fraction relative to galaxies with a similar stellar mass of $\sim 10^{10.4}$ M$_\odot$, and exhibits a high specific star formation rate.

## 1 INTRODUCTION

Collisional ring galaxies are peculiar systems that highlight the important role of gravitational interactions in the formation and evolution of galaxies (Lynds & Toomre 1976; Theys & Spiegel 1977; Struck-Marcell & Lotan 1990). These systems usually result from the passage of a compact companion galaxy (the 'intruder') through the disc of a more massive spiral galaxy (the 'target'). Such drop-through collisions generate a density wave that radially carries/transports the gas and stars into a ring morphology throughout the disc of the target galaxy (Appleton & Struck-Marcell 1996; Mayya et al. 2005; Wong et al. 2006; Romano, Mayya & Vorobyov 2008; Fogarty et al. 2011; Parker et al. 2015; Conn et al. 2016). The density of the ring and the perturbation in the target's velocity field depend on the mass of the intruder; that is, high-mass intruders produce denser rings, faster expansions and multiple rings in some cases (Gerber, Lamb & Balsara 1996). Depending on the impact parameter, the target's nucleus may be displaced from its dynamical centre, and the entire system (ring and nucleus) may be slightly dislocated from the original plane of the disc (Mapelli & Mayer 2012). In some cases the nucleus will be completely disrupted by the collision, resulting in a centrally smoothed ring with no apparent nucleus (Madore, Nelson & Petrillo 2009). Furthermore, Fiacconi et al. (2012) showed that certain impact parameters and collision angles can induce warps in the disc and affect the morphology and the star formation history of the target galaxy. Because the formation mechanism of collisional ring galaxies is well constrained, these systems can often be considered as galaxy-scale perturbation experiments that facilitate the study of extreme modes of interaction-triggered star formation and feedback processes.

Simulations and observations of the H I gas morphology and kinematics of ring galaxies can provide important constraints on their formation mechanism, the mass ratios, the time-scale and the dynamics of the collision. Here, we study the H I gas morphology and kinematics of the collisional ring galaxy NGC 922 using new H I observations from the Australia Telescope Compact Array (ATCA). This galaxy has been observed in the H$\alpha$ and $R$ band by the Survey for Ionization in Neutral Gas Galaxies (SINGG; Meurer et al. 2006), in the far-ultraviolet (FUV) and the near-ultraviolet (NUV) bands by the Survey of Ultraviolet emission of Neutral Gas Galaxies (SUNGG), and using the *Hubble Space Telescope* (*HST*) Wide Field Planetary Camera 2 (WFPC2) (Pellerin et al. 2010, hereafter P10). Wong et al. (2006, hereafter W06) showed that the stellar plume observed in NGC 922 and the partial ring morphology seen in its disc are characteristics of a drop-through collision of a companion dwarf galaxy. Using numerical simulations, W06 were able to reconstruct all the major features of this system, including the radius of the ring and the projected physical distance between NGC 922 and its companion (hereafter S2). NGC 922 lies at a systemic velocity of $cz \approx 3077$ km s$^{-1}$ and has a star formation rate (SFR) of $\sim 8$ M$_\odot$ yr$^{-1}$, while S2 has a SFR of $\sim 0.47$ M$_\odot$ yr$^{-1}$. Table 1 presents the general physical properties of NGC 922 and S2.

[*] E-mail: ahmed.elagali@icrar.org

**Table 1.** General properties of NGC 922 and the intruder S2 from Wong et al. (2006) and Pellerin et al. (2010).

| Property | NGC 922 | S2 |
|---|---|---|
| Right ascension (J2000) | 02:25:04 | 02:24:30 |
| Declination (J2000) | −24:47:17 | −24:44:44 |
| $cz$ (km s$^{-1}$) | 3077 | 3162 |
| $L_R$ (L$_\odot$) | $3.7 \times 10^{10}$ | $1.4 \times 10^9$ |
| H $\alpha$ equivalent width (Å) | $77 \pm 3$ | $43 \pm 5$ |
| $(FUV - NUV)_0$ (ABmag) | −0.09 | −0.08 |
| $(NUV - R)_0$ (ABmag) | 1.52 | 1.32 |
| Stellar mass (M$_\odot$) | $2.8 \times 10^{10}$ | $1.8 \times 10^9$ |
| SFR$_{H\alpha}$(M$_\odot$ yr$^{-1}$) | $8.20 \pm 0.32$ | $0.26 \pm 0.03$ |
| SFR$_{FUV}$(M$_\odot$ yr$^{-1}$) | $7.04 \pm 0.02$ | $0.47 \pm 0.02$ |
| $Z$ (Z$_\odot$) | 0.5 − 1.0 | 0.3 − 0.5 |
| Distance (Mpc) | 43.7 | 44.9 |
| Physical diameter [$2a$] (kpc) | 30 | 6.5 |
| Age of the ring (Myr) | 300 | - |
| Angular separation (arcmin) | 8 | 8 |
| Projected physical distance (kpc) | 102 | 102 |

Fig. 1(a) shows the SINGG three-colour RGB composite image of NGC 922 and its 'intruder' S2, with the H $\alpha$, H $\alpha+$ continuum and *R*-band images representing the red (R), green (G) and the blue (B), respectively. The inset is a deep grey-scale image of NGC 922 created from the UK Schmidt plates, courtesy of David Malin.[1] The stellar plume seen in this grey-scale image extends towards the direction of the companion dwarf galaxy S2, and is one of the reasons to favour an interaction history to explain the peculiar morphology of NGC 922, as opposed to secular processes. Figs 1(b) and (c) are zoom-in images of NGC 922 using a three-colour RGB image composited similarly to that in the upper panel and the SINGG *R*-band filter image, respectively. The H $\alpha$ emission regions in this galaxy are distributed in a partial ring morphology (C-shape) and along the bar with a radius of 15 kpc. The bar and the bulge of NGC 922 are off-centre towards the northern part of the disc. The majority of the prominent H $\alpha$ emission and star-forming knots are on the eastern side of the ring, while the western side hosts regions with a low SFR and weak H $\alpha$ emission. Notable H $\alpha$ emission is also present on the southern edge of this galaxy, but is disconnected from the C-shape morphology and apparently is associated with giant/supergiant H II regions. P10 found, using the *HST*/WFPC2 images, that almost 70 per cent of the stellar clusters in NGC 922 are very young, with ages of less than 7 Myr. The majority of these young clusters are located in the partial ring of NGC 922 (see Fig. 1c), which agrees well with the H $\alpha$ distribution. The remaining clusters have ages ranging between 7–50 Myr and

$\leq$100 Myr, and are located mainly in the nucleus but with some distributed along the bar.

Using these new H I observations, we will study here the consequences of the drop-through collision on the atomic gas morphology and kinematics of NGC 922 and compare it with other nearby collisional systems. This comparison will shed light on the full interaction history between NGC 922 and S2, which will enhance our understanding of the peculiar morphology of this galaxy. We will also explore multiwavelength observations of NGC 922 and compare its properties with those of nearby star-forming galaxies, for instance in terms of the H I gas mass fraction, specific SFR and dust temperature. This paper is organized as follows. In Section 2, we describe our ATCA H I observations, data reduction and analysis. In Section 3 and 4, we present our results and discussion, with a focus on the gas morphology and kinematics. In Section 5, we summarize our main findings. For consistency with W06 and P10, we have adopted a $\Lambda$CDM cosmology with the following parameters: $\Omega_m = 0.27$, $\Omega_\Lambda = 0.73$, and a Hubble constant of $H_0 = 70.3$ km s$^{-1}$ Mpc$^{-1}$ (Komatsu et al. 2011).

## 2 DATA

### 2.1 Observations

We observed NGC 922 and S2 using the ATCA radio interferometer over 4 d between 2013 October and 2014 January. The ATCA is a set of six 22-m-diameter radio dishes located at $-30°$ latitude, and situated on an east–west track with a longest baseline of 6 km. To optimise between the surface brightness sensitivity and the angular resolution of our observations, three ATCA array configurations were used, namely the H214, 750B and 1.5B configurations. We combined our data with additional ATCA archival observations obtained on 2005 November 26, using the 1.5B configuration. The baseline range for these arrays is between 34 and 4500 m. The observations were centred at the redshifted H I line frequency of the targets at 1396 MHz, with a total on-source integration time of 29.5 h. We used the Compact Array Broad-band Backend (CABB; Wilson et al. 2011) of the telescope in the mode that provides a 64-MHz bandwidth over 2048 velocity channels, resulting in a velocity resolution of 6 km s$^{-1}$. Tables 2 and 3 present a summary of our ATCA H I observations and results.

### 2.2 Data reduction and analysis

We used the Australia Telescope National Facility (ATNF) version of the MIRIAD (Multichannel Image Reconstruction, Image Analysis and Display) package (Sault, Teuben & Wright 1995) to process and calibrate the spectral line data. The flux and the bandpass were calibrated using the radio source $1934-638$ (primary calibrator). To improve the calibration precision, the phase changes were monitored each hour throughout the observations using the secondary calibrator $0237-233$, which has a flux density of 6.09 Jy relative to the flux density of the primary calibrator, $S_{1396\,MHz} = 14.9$ Jy. Both the antenna gains and the bandpass calibration were applied to the targets to form Stokes I. We used the UVLIN task in MIRIAD (Cornwell, Uson & Haddad 1992) to generate a fit through the line-free channels of the observed visibilities and subtracted this fit from all the frequency channels. The continuum-subtracted data were transformed into a map using the MIRIAD INVERT task. The H I line cube was made after imaging with robust weighting (Robust = +0.4), which gave optimal sidelobe suppression for the map

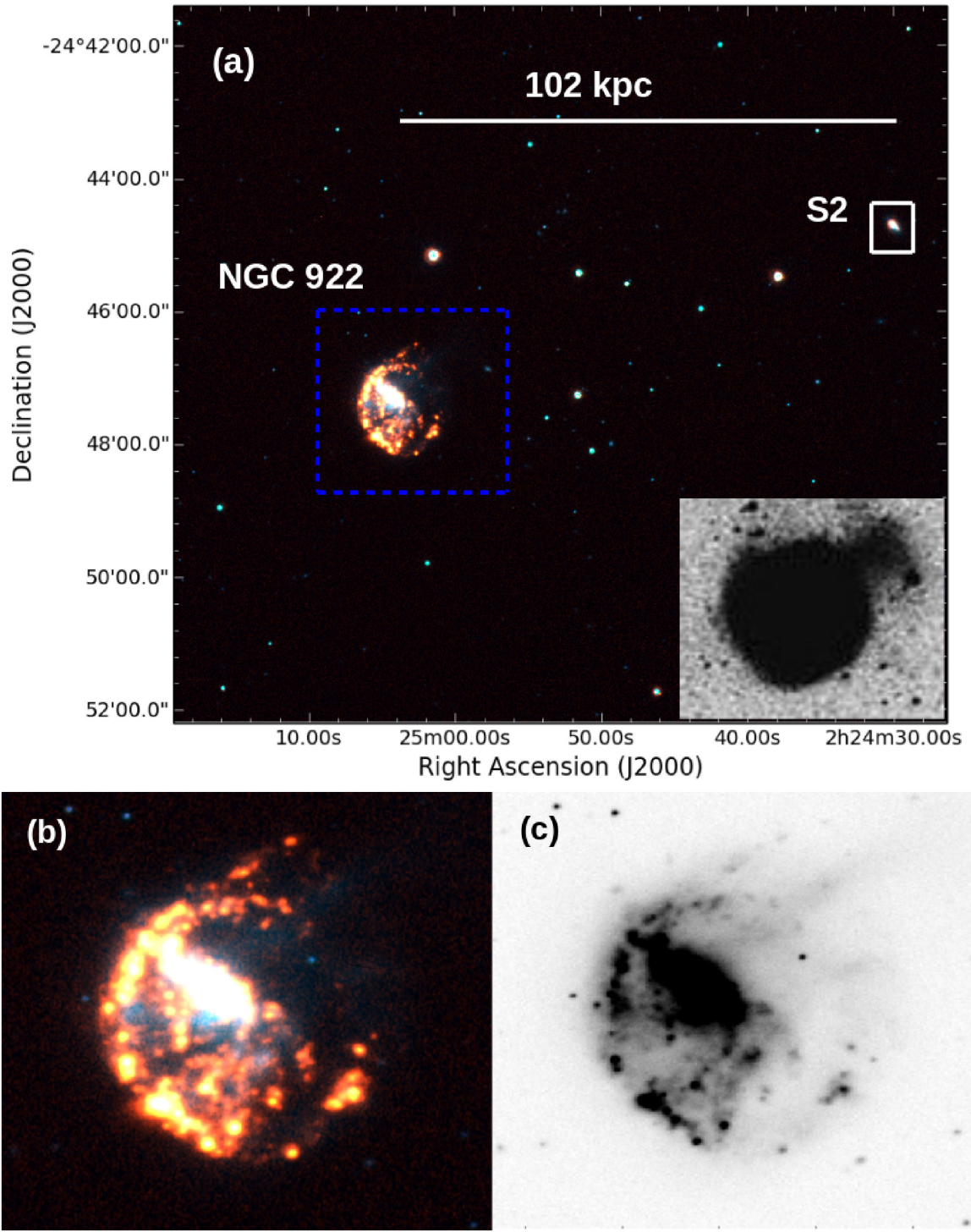

**Figure 1.** (a) The SINGG three-colour RGB image of the collisional ring galaxy NGC 922 and its 'intruder' S2. In this colour composite, the H $\alpha$, H $\alpha$+ continuum, and *R*-band images are represented in red (R), green (G) and blue (B), respectively. The H $\alpha$ emission regions in this galaxy are distributed in a C-shape morphology, induced by the passage of S2 through the disc of NGC 922, and along the bar with a radius of 15 kpc. The projected physical distance between NGC 922 and S2 is 102 kpc (Wong et al. 2006). The inset is a zoom-in deep grey-scale image of NGC 922 created from the UK Schmidt plates. Panels (b) and (c) are zoom-in images of NGC 922 using a three-colour RGB image composited similarly to the upper panel and the SINGG *R*-band filter image, respectively. The RGB images are made using a square root (sqrt) colour stretch to highlight regions of low star formation, while the *R*-band image is made using a linear colour stretch.

and intermediate weighting between uniform and natural. The cube was then cleaned and smoothed to obtain a velocity resolution of 10 km s⁻¹. Thereafter, we applied primary beam corrections to the restored cube using the LINMOS task. The restored beam has a size of $\theta_{\rm FWHM} = 40 \,\rm arcsec \times 1 \,arcsec$ with a position angle (PA) = 0. The cube has an rms noise of 1.0 mJy beam⁻¹ per channel, close to the theoretical rms noise of our observations of 0.83 mJy beam⁻¹ per channel. We discarded visibility data from the antenna 6

**Table 2.** ATCA H I observations: instrumental parameters.

| Parameter | Array configuration | | | |
|---|---|---|---|---|
| | 1.5B | 750B | H214 | 1.5C |
| Observation dates | 2014 January 15-16 | 2013 December 10 | 2013 October 17 | 2005 November 26 |
| On-source integration time (hr) | 12.3 | 10.2 | 4.0 | 3.0 |
| Shortest baseline (m) | 31 | 61 | 92 | 77 |
| Longest baseline (m) | 4301 | 4500 | 4500 | 4500 |
| Central frequency (MHz) | 1396 | 1396 | 1396 | 1396 |
| Bandwidth (MHz) | 64 | 64 | 64 | 16 |

**Table 3.** ATCA H I observation results.

| Parameter | Value |
|---|---|
| rms noise (Jy beam$^{-1}$) | $1.0 \times 10^{-3}$ |
| $3\sigma$ column density (cm$^{-2}$) | $4.59 \times 10^{19}$ |
| Synthesized beam size (arcsec × arcsec) | 40 × 18 |
| PA (°) | 0 |
| Channel width (km s$^{-1}$) | 10.0 |
| NGC 922 H I total flux (Jy km s$^{-1}$) | 24.7 ± 1.0 |
| NGC 922 peak flux density (Jy) | 0.13 ± 0.03 |
| NGC 922 H I mass ($M_\odot$) | $(1.10 \pm 0.04) \times 10^{10}$ |
| NGC 922 line width (km s$^{-1}$) | 188 ± 10 |

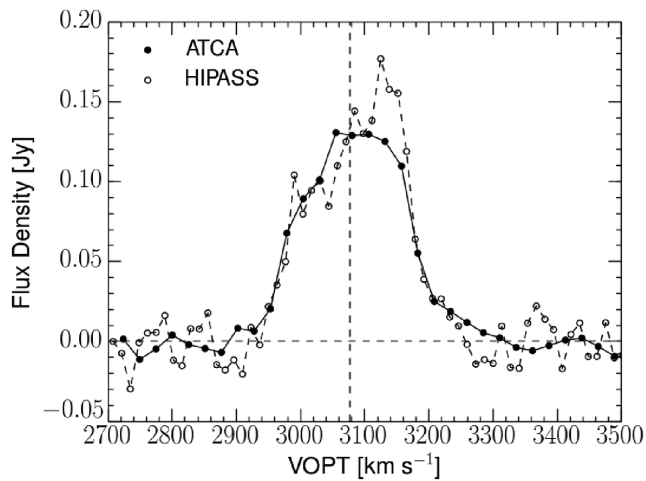

**Figure 2.** The global H I line profile of NGC 922 from our ATCA observations (filled circles) and from the H I Parkes All-Sky Survey (open circles). The integrated H I flux density of NGC 922 from our ATCA observations (24.7 ± 1.0 Jy km s$^{-1}$) is within the error of the flux values reported in the HIPASS galaxy catalogue (26.9 ± 2.4 Jy km s$^{-1}$; Meyer et al. 2004) and in the Brightest HIPASS Galaxies Catalogue (28.1 ± 4.1 Jy km s$^{-1}$; Koribalski et al. 2004). The surface bright-

The vertical line delimits the radio systemic velocity of NGC 922 $V_{sys} = 3077$ km s$^{-1}$, while the horizontal line is the fitting baseline.

baseline to improve our surface brightness sensitivity towards extended and diffuse H I emission.

## 3 RESULTS

### 3.1 H I distribution of NGC 922

Fig. 2 shows the global H I line profile of NGC 922 from our ATCA observations as filled circles and the data from the H I Parkes All-Sky Survey (HIPASS; Barnes et al. 2001) as open circles. The integrated H I flux density of NGC 922 from our ATCA observations (24.7 ± 1.0 Jy km s$^{-1}$) is within the error of the flux values reported in the HIPASS galaxy catalogue (26.9 ± 2.4 Jy km s$^{-1}$; Meyer et al. 2004) and in the Brightest HIPASS Galaxies Catalogue (28.1 ± 4.1 Jy km s$^{-1}$; Koribalski et al. 2004). The surface bright-ness sensitivity of HIPASS observations is higher than the surface brightness sensitivity for our 29.5 h of ATCA observations : the $3\sigma$ column density limit of HIPASS is $1.02 \times 10^{19}$ cm$^{-2}$ (Koribalski et al. 2004), which is probably the reason for the difference in the measured flux between the two observations. The total H I mass of NGC 922 derived from our combined ATCA observations is $1.1 \times 10^{10}$ M$_\odot$, assuming a distance of 43 Mpc (W06). The difference between our new observations and those from HIPASS suggests that our new observations may be missing flux (8–12 per cent) on the receding side of the spectrum (Fig. 2). This may also explain the position offset of the H I centre in our ATCA map with respect to the HIPASS H I map centre. The position of the ATCA map centre makes more sense, as it is located exactly at the nucleus of NGC 922, while the position of the HIPASS map centre is denoted by the filled black diamond shown in Fig. 3. The offset in right ascension between the two centres is $\Delta\alpha = 1.5 \pm 0.6$ arcmin, which is close to the positional accuracy of HIPASS observations ($\Delta\alpha = 1.0$ arcmin) at the 99 per cent completeness limit (Zwaan et al. 2004). Furthermore, the integrated intensity maps over the velocity channels between 3100 and 3170 km s$^{-1}$ for the ATCA and HIPASS observations have different centroids. The positions of these centroids are shown by the green circle (ATCA map) and blue star (HIPASS map) in Fig. 3. The HIPASS centroid (at velocities higher than 3100 km s$^{-1}$) is extended towards S2 (see Fig. 1a). Hence, the 'missing' H I flux from the ATCA observations may be in the form of diffuse gas in the direction of S2 that has been resolved out. In W06's simulations, an H I tail appears to coexist with the stellar plume, along the direction of the collision and towards the intruder S2, at earlier snapshots of the simulation (∼100 to 200 Myr after the collision) but disappears 100 Myr later.

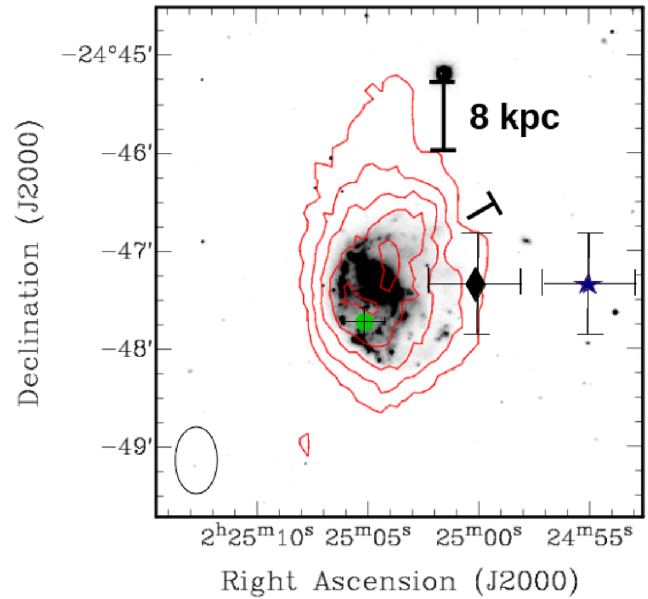

**Figure 3.** The ATCA integrated H I column density distribution of NGC 922 superimposed on the optical *R*-band image obtained from the SINGG survey. The column density contours are at (2.3, 4.5, 9.0, 13.5, 18.0) × 10$^{20}$ cm$^{-2}$, which is probably the reason for the difference. The H I centre of this ATCA map is different from the centre of a previous H I map of NGC 922 obtained using the Parkes telescope as part of the HIPASS survey. The ATCA map centre is located exactly at the nucleus of NGC 922, while the HIPASS map centre is denoted by the filled black diamond; the offset between the two centres is $\Delta\alpha = 1.5 \pm 0.6$ arcmin. The blue star shows the HIPASS intensity map centroid over the missing flux velocity channels (3100–3170 km s$^{-1}$), while the green circle shows the ATCA centroid over the same velocity channels. The 'T' delimits the location of the stellar plume emerging from NGC 922 towards S2. The error bars are the position uncertainties associated with each measurement.

Fig. 3 shows the ATCA H I column density distribution of NGC 922 overlaid on the SINGG *R*-band image. The H I gas is slightly offset to the north of this galaxy with a tail also pointing northwards. Furthermore, the H I is co-located with regions of high rates of star formation and young (≤7 Myr) stellar clusters (P10). Even though the surface brightness sensitivity of our observations is sufficient to detect relatively low H I column densities, we have not detected any H I emission (above our rms noise) from the intruder S2. The upper mass limit of S2 based on our rms noise ($3\sigma$) and assuming three velocity channel widths (30 km s$^{-1}$) is <4 × 10$^7$ M$_\odot$. Even though S2 is an H I-deficient galaxy, this does not undermine the identity of S2 as the possible intruder. This is because the SINGG narrow-band H α image of this field showed that S2 is the only companion of this ring galaxy, and hence the only possible intruder. We note that the next closest candidate to NGC 922 is AM 0223-245 NED03, which is 160 kpc (projected physical distance) from NGC 922 and has a recessional velocity of 10 216 km s$^{-1}$. Furthermore, in most of the studied ring galaxies, H I is detected in the intruder and/or there is an H I tail in its direction (for example Higdon 1996). In our case, we find an H I tail that points in a different direction with respect to the direction of the intruder S2. The H I tail extends north of the disc to ∼0.7 arcmin (8 kpc), after convolving the optical image to the beam size of our H I map (see Fig. 3). This is inconsistent with the simple collisional scenario proposed for NGC 922 (by W06), because the H I tail (north) and the stellar plume (marked by the rotated 'T' in Fig. 3) are in different directions. Instead, this is suggestive of a more complex formation history for NGC 922. For instance, a tidal interaction might have induced the formation of this H I tail before or after the drop-through collision between NGC 922 and S2.

Fig. 4 presents the channel maps of NGC 922 from our ATCA observations. The H I gas is more extended to the north, which suggests that the passage of S2 through the disc of NGC 922 was off-centre and with a large impact parameter. If the companion S2 plunged through the centre of NGC 922 (a 'bullseye' collision), the H I gas would have expanded uniformly in all directions (see fig. 2 in Fiacconi et al. 2012). This agrees with the W06 simulations, in which the C-shape morphology of NGC 922 was reproduced by a high-speed (481 km s$^{-1}$) off-centre passage of the companion S2 through the NGC 922 disc with an impact parameter of 6.7 kpc (see fig. 3 in W06). However, the W06 simulations fail to account for the aforementioned H I tail that points north. It is important to note that, in these simulations, NGC 922 was represented by a dark matter halo and a thin exponential disc with a gas mass fraction of 0.2, and S2 was represented as a point mass based on the data then available. Moreover, W06 assume the simplest drop-through ring formation scenario, and hence it is not surprising that the simulation underestimates the extent of the gas seen in Fig. 3. Likewise, the inconsistencies between the observed and the modelled gas morphologies are not surprising.

## 3.2 The kinematics and velocity fields of NGC 922

In order to understand the kinematics of NGC 922, we derived the rotation curve of this galaxy using a tilted-ring model (Rogstad, Lockhart & Wright 1974). In practice, the tilted-ring model fits the following equation to the observed velocity field:

$$V(x, y) = V_{sys} + V_{rot}\,\sin i \cos\theta, \qquad (1)$$

where $V(x, y)$ is the line-of-sight velocity at a sky position $(x, y)$ in the ring and a distance $R$ from the rotation centre, $V_{sys}$ is the systemic velocity, $V_{rot}$ is the circular velocity, and $\theta$ is the azimuthal angle from the major axis in the galaxy plane. This angle is related to the inclination ($i$) and the position angle (*PA*) of the disc.

We fitted a tilted-ring model using the Groningen Image Processing System (GIPSY; Allen, Ekers & Terlouw 1985; van der Hulst et al. 1992) ROTCUR task (Begeman 1989, 1987) to derive the rotation curve of NGC 922. It is difficult to extract a unique solution for all the parameters in equation (1) because our H I observations have only four synthesized beams across the major axis. Therefore, in order to model the disc of NGC 922, we ran the task in an iterative fashion and used a $|\cos\theta|$ weighting function to maximize the contribution of the velocity points close to the major axis. First, we fitted the dynamical centre and the systemic velocity $V_{sys}$ simultaneously by keeping *PA* (172°) and $i$ (52°) fixed to their optical values. Then, keeping the dynamical centre and the systemic velocity fixed, we fitted the inclination $i$ and the position angle *PA* simultaneously, because they are correlated. Thereafter, the *PA* and $i$ values were smoothed using a boxcar function and fixed while running the task for the final time to estimate a unique solution for $V_{rot}$. The ROTCUR tilted-ring fit corresponds to a systemic velocity of $V_{sys} = 3074 \pm 7$ km s$^{-1}$ and a dynamical centre that coincides with the *HST* isophotal centre. The best-fitting values for the position angle and the inclination are $PA = 157° \pm 5°$ and $i = 54° \pm 4°$, respectively. While the inclination derived from the kinematic model is within the error of the optical value (52°), the position angle obtained is smaller than optical position angle derived from the ellipse fit to the optical *R*-band image in Fig. 3 (172°). The difference between the PA of the H I gas and the stars is probably a result of the complicated interaction history between NGC 922 and S2. The gaseous disc of NGC 922 seems to be more affected by the interaction (as shown in Section 3.1) than the stellar counterpart of the galaxy. This is in agreement with previous H I studies of interacting galaxies, in which the H I discs are more sensitive to external influences than their optical counterparts (Michel-Dansac et al. 2010; Braun & Thilker 2004; Yun, Ho & Lo 1994).

Fig. 5 shows the ROTCUR solutions (approaching, receding, both sides) for *PA*, $i$ and the rotation velocity of NGC 922. The variation in *PA* and $i$ outside the optical radius suggests that the H I disc of NGC 922 is mildly warped. If this H I disc was not warped then the fitted position angle and inclination would be similar throughout the various ring radii. In our case, however, the difference in *PA* and $i$ between the inner ($r = 6$ kpc) and the outer ($r = 25$ kpc) ring is over 20°. The green line in the upper panels of Fig. 5 shows the *PA* and $i$ values used to derive the overall rotation curve of NGC 922. The rotation curve derived using ROTCUR has many factors that dominate its uncertainties, such as the non-circular motion of the observed gas, and the angular resolution of our observations. To highlight these systematic uncertainties, we derive the rotation curve for the approaching (filled stars in Fig. 5c) and the receding

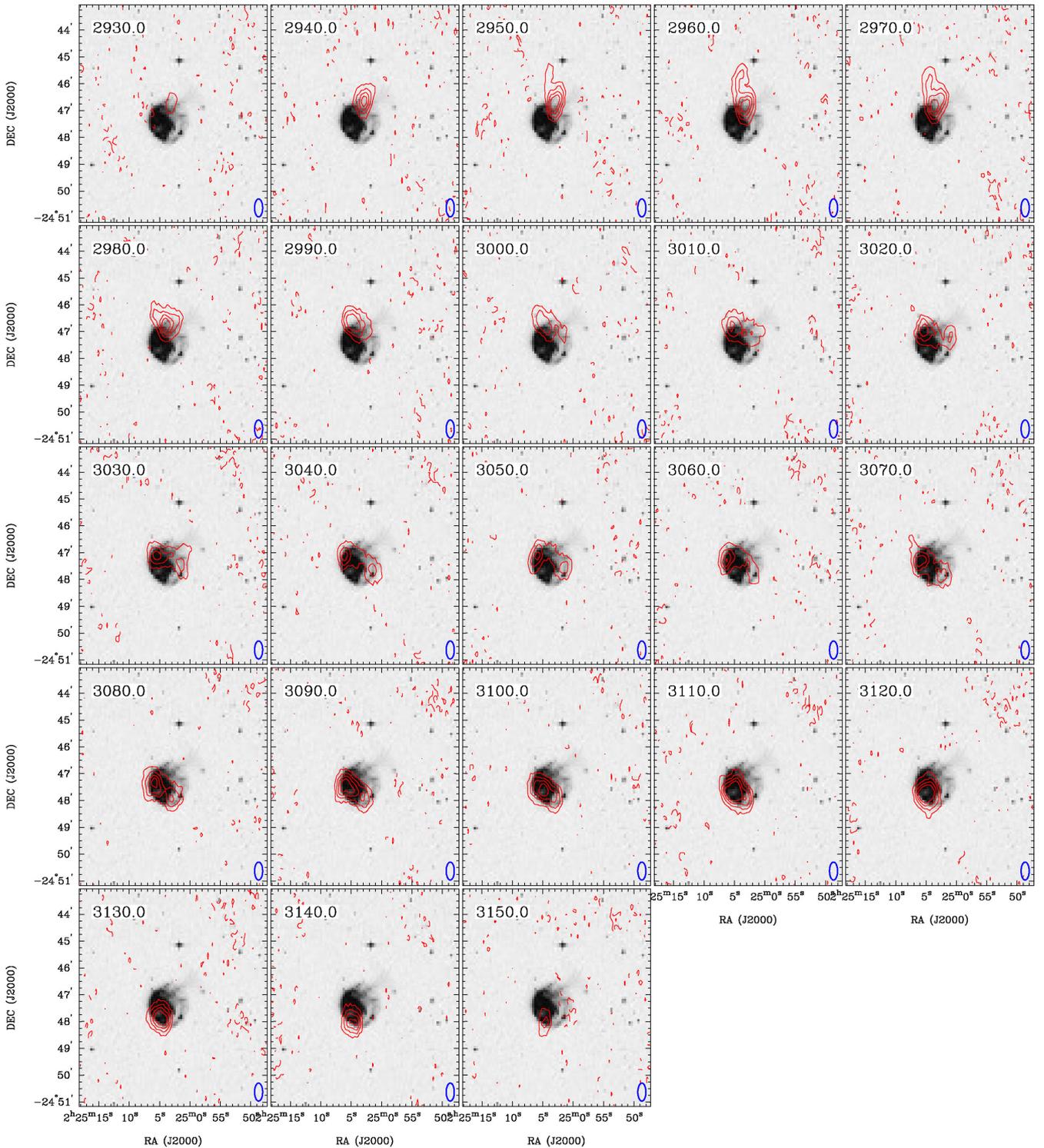

**Figure 4.** H I channel maps of NGC 922 as obtained from the ATCA observations using robust weighting (Robust = +0.4) superimposed on the SINGG *R*-band image. The velocities of these maps range from 2930 to 3150 km s$^{-1}$, with a step size of 10 km s$^{-1}$, and are shown in the top-left of each panel. The contours are −3, −1, 1, 3, 5, 7, 9 times the 1 σ noise (1 mJy beam$^{-1}$) in a single velocity channel (10 km s$^{-1}$). The synthesized beam size is $\theta_{FWHM}$ = 40 arcsec × 18 arcsec, shown as a blue ellipse in the bottom-right of each panel, and has a position angle (PA) of 0°.

(open circles in Fig. 5c) sides, independently, and compare the two with the overall rotation curve (filled blue circles in Fig. 5d) of this galaxy. In contrast to the overall rotation curve and the approaching solutions, the receding solutions reach a maximum velocity and then decrease at larger radii. However, the solution for the approaching side differs in the shape and the position of the maximum velocity compared with that of the receding side; the approaching side has its maximum farther out ($r$ = 22 kpc) with respect to the receding side ($r$ = 18 kpc). The difference between the approaching/receding sides indicates an asymmetry in the gas kinematics and an overall systematic uncertainty of up to $\pm 10\,\mathrm{km\,s^{-1}}$. The uncertainties are higher than $\pm 10\,\mathrm{km\,s^{-1}}$ in some parts of the NGC 922 disc (inner and outer rings). The overall rotation velocity of this galaxy reaches a maximum of $125 \pm 6\,\mathrm{km\,s^{-1}}$ at the outermost part of the H I disc. This rotation velocity corresponds to, assuming a spherically symmetric mass distribution, a total mass enclosed within the outermost data point of $M_{\mathrm{tot}} = (9.1 \pm 0.8) \times 10^{10}\,M_\odot$. Thus the H I mass fraction ($M_{\mathrm{H\,I}}/M_{\mathrm{tot}}$) in NGC 922 is approximately 0.11, and the stellar mass fraction ($M_*/M_{\mathrm{tot}}$) is 0.30. This dynamical mass corresponds to an escape velocity ($v_{\mathrm{esc}}$) of $170 \pm 10\,\mathrm{km\,s^{-1}}$ (assuming a radius of 25 kpc), while the predicted relative velocity of S2 from the W06 numerical simulation is $v = 203\,\mathrm{km\,s^{-1}}$. This suggests that S2 is unlikely to be bound to the gravitational potential of NGC 922.

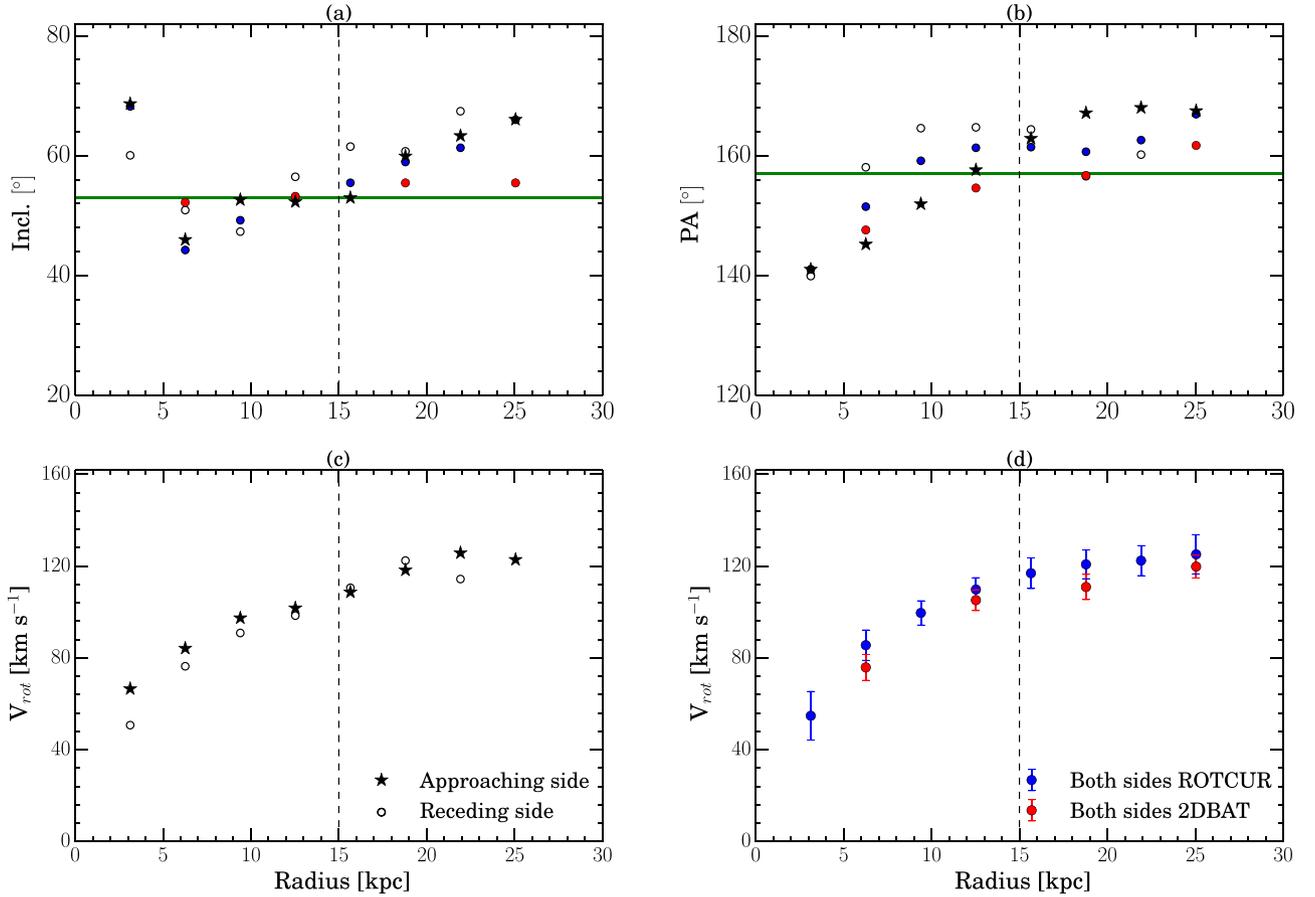

**Figure 5.** Panels (a) and (b) show the tilted-ring model solutions (approaching, receding, both sides) for the inclination *i* and the position angle *PA* of NGC 922 using ROTCUR, with filled stars, open circles and filled blue circles. The red filled circles in the upper two panels show the tilted-ring model solution for both sides using 2DBAT. Panel (c) shows the rotation curve for the approaching (filled stars) and receding (open circles) sides using ROTCUR. Panel (d) shows the overall rotation curve of NGC 922 derived using ROTCUR with filled blue circles and using 2DBAT with filled red circles. The green line in the upper two panels shows the adopted position angle ($PA = 157°$) and inclination ($i = 54°$) used to derive the overall rotation curve in (d). The vertical dashed line indicates the optical radius of this galaxy.

The observed H I intensity-weighted mean velocity field of NGC 922 is shown in Fig. 6(a); the model velocity field derived using ROTCUR is shown in Fig. 6(b); and the residual map between the two velocity fields is shown in Fig. 6(c). The contours in Fig. 6(a) and (b) are centred on the systemic velocity of NGC 922 ($3074\,\mathrm{km\,s^{-1}}$) and separated at intervals of $20\,\mathrm{km\,s^{-1}}$. Ideally, for an unperturbed H I disc the majority of the H I velocity residuals should lie within $\pm 10\,\mathrm{km\,s^{-1}}$. In our case, however, there are large velocity residuals in certain regions, mostly in the inner and the outer disc, in excess of $\pm 40\,\mathrm{km\,s^{-1}}$. This is almost 30 per cent of the maximum rotation velocity of the outer disc. We attribute this discrepancy in the velocity of the H I disc to the variations in the position angle and the inclination with radius. Such variations in *PA* and *i* are indicative of warps in the H I disc (especially beyond the optical radius) and/or of substantial distortions caused by the collision with the intruder. Furthermore, evidence for a warped H I disc is noticeable in the observed velocity field of NGC 922. The radial variation of the position angle is prominent in the approaching side of the observed velocity field and especially on the western side of the H I disc. As an additional check, we used a different package to fit the rotation curve of NGC 922, namely the 2D Bayesian Automated Tilted-ring fitter of disc galaxies in large H I galaxy surveys (2DBAT; Oh et al. 2018). This new software uses the Bayesian Markov chain Monte Carlo (MCMC) technique to optimise the tilted-ring fit as opposed to the traditional chi-square minimization procedure in ROTCUR. Hence, solutions from 2DBAT are less susceptible to being trapped in local minima. The rotation curve and the model velocity field of NGC 922 derived using 2DBAT (shown by the filled red circles in Fig. 5) are consistent with those derived using ROTCUR and presented in this section.

Previous studies of collisional ring galaxies have found signatures of expansion in their rings. The expansion velocity can reach up to $113\,\mathrm{km\,s^{-1}}$ for recently formed rings ($\sim 50\,\mathrm{Myr}$), as in the

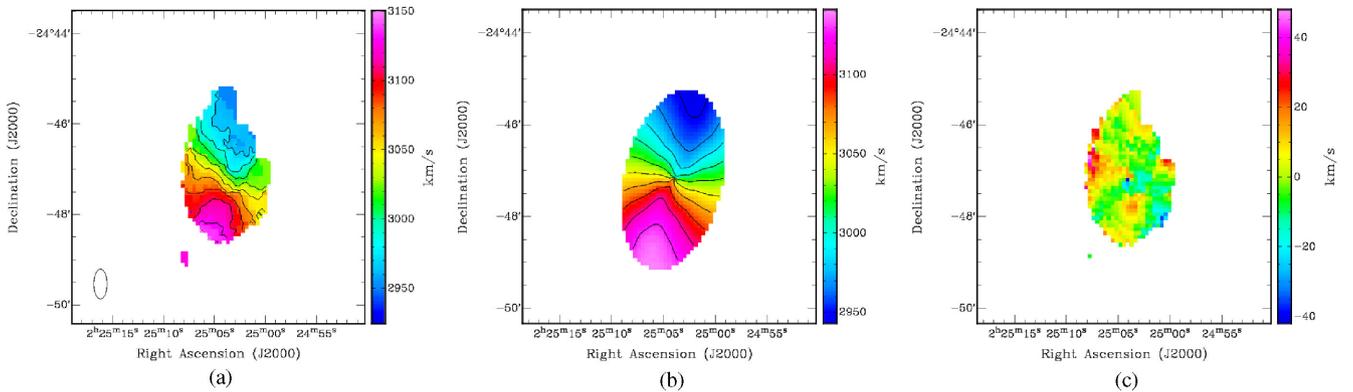

**Figure 6.** Panel (a) shows the velocity field of NGC 922 based on our H I moment map, while panel (b) shows the velocity field derived from the tilted-ring ROTCUR fit estimated assuming $V_{exp} = 0 \, km \, s^{-1}$. The contours in (a) and (b) are centred on the systemic velocity of NGC 922 (3074 km s$^{-1}$) and separated at intervals of 20 km s$^{-1}$. Panel (c) shows the difference between the observed velocity and the model. The ellipse in the bottom left corner of (a) shows the synthesized beam of the ATCA observations.

Arp 147 ring galaxy (Fogarty et al. 2011), and about ~50 km s$^{-1}$ for older rings (~200 Myr), as in the Cartwheel galaxy (Higdon 1996). Hence, we investigated this possibility in NGC 922 by adding an expansion velocity ($V_{exp}$) term to equation (1) and running the ROTCUR task again. We followed the same procedures as described earlier and fitted for the rotation and the expansion velocities simultaneously, while keeping the dynamical centre, $V_{sys}$, PA and $i$ at the best-fitting values. We found the expansion velocity ($V_{exp}$) of NGC 922 derived from the tilted-ring model to be approximately 0 km s$^{-1}$ and to have no significant effect on the modelled rotation velocity of the disc. The zero expansion velocity in NGC 922 is different from the case of other collisional ring galaxies, but not surprising considering the time elapsed since the collision ($t \sim 250$ Myr ; W06) and the relatively low mass of the intruder S2.

## 4 DISCUSSION

### 4.1 Gas morphology and kinematics of NGC 922

Simulations of collisional ring galaxies show that a wide range of morphological features can be formed for different combinations of the impact parameter and the intruder velocity, for example single rings (Arp 143, Kathryn's wheel), double rings (Cartwheel, Lindsay-Shapley Ring), half rings (Arp 107, Arp 148), as well as empty rings (VII Zw466, Arp 147). In NGC 922, the collision was off-centre in a northerly direction. The high-velocity passage of the intruder (~481 km s$^{-1}$) and the high impact parameter (6.7 kpc) have had a significant influence on the morphology of NGC 922 and are probably the reason for its stellar C-shape morphology (W06) and the extended H I distribution to the north.

Although there is a plume of old stars (visible in the SINGG *R*-band image) emerging from NGC 922 towards the intruder (north-west direction), we find no H I tail that is coincidental with this stellar plume. This is probably because of the small collision angle between NGC 922 and S2 (inferred from W06's simulation); the observed morphology of NGC 922 is best matched assuming a collision angle of $\theta \sim 0°$. Simulations of collisional ring galaxies show that the collision angle plays an essential role in forming the H I tails observed between ring galaxies and their intruders. These tails are more pronounced in collisional systems formed with acute collision angles $\theta > 0°$ (see section 3 in Fiacconi et al. 2012). The case of

NGC 922 is therefore different from many observed collisional ring galaxies that possess H I tails towards their intruders.

Even though there is no H I tail that coexists with the stellar plume and points towards S2, we find an H I tail that points towards the north of the disc and extends to 8 kpc (evident at velocities between 2950 and 2980 km s$^{-1}$ in Fig. 4). NGC 922 is comparable to the umbrella galaxy, NGC 4651, which experienced a satellite dwarf plunging through its disc creating a stellar plume to the east (the umbrella-like feature) and an H I tail that points to the western part of its disc (Chung et al. 2009; Foster et al. 2014). One possible scenario that explains the offset H I tail is that an additional tidal interaction between NGC 922 and S2 occurred before or after the drop-through collision. However, we caution that this scenario is still inconclusive. It is likely the Integral Field Unit (IFU) observations will provide some stellar kinematics of NGC 922 that may constrain future simulations and reveal the true formation history of this galaxy.

Many collisional ring galaxies exhibit evidence of warped outer discs, for example UGC 7069 (Ghosh & Mapelli 2008) and AM1354-250 (Conn et al. 2016). The warps result from the net torque induced by the intruder and depend on the geometry of the collision. After the interaction, the target galaxy will be displaced from its original position and the ring will develop in a plane tilted to the original plane of the disc, which will generate warps. In NGC 922, the high speed of the intruder and the impact parameter play a dominant role in inducing warps, because the collision angle is very small. The gas warps are evident in the velocity field of NGC 922; the H I disc apparently is more warped on the approaching side starting from 3060 km s$^{-1}$, and along the western side of the disc. The discrepancy seen in the residual field can thus be attributed to the warp/distortion features apparent in the H I disc.

### 4.2 Gas fraction and star formation in NGC 922

In order to investigate the H I gas fraction in NGC 922, we compared the results from our observations with the standard H I gas fraction scaling relationships found for two samples of nearby star-forming galaxies. The first sample is presented in Cortese et al. (2011) and was used to investigate the effects of a dense environment on the gas fraction of galaxies. Here, we compare NGC 922 with the 'H I-normal' galaxies of this sample (135 galaxies). These are unperturbed/isolated galaxies with an H I-deficiency[2] < 0.5, within a mass range of $10^9 \lesssim M_* \lesssim 10^{11}$ and within a distance range of $15 \lesssim d \lesssim 25$ Mpc. The second sample comprises a volume-limited sample of 25 000 galaxies from the Sloan Digital Sky Survey, with H I spectra from the Arecibo Legacy Fast (ALFALFA) survey (Giovanelli et al. 2005). The H I masses were determined using stacking techniques and subsequently used to investigate the dependence of the H I gas content on galaxy properties, such as the surface mass, the specific star formation rate (sSFR) and the stellar surface density (Brown et al. 2015). The sample lies within a stellar mass range of $10^9 \lesssim M_* \lesssim 10^{11.5}$ and with a distance within the range $80 \lesssim d \lesssim 200$ Mpc, which is greater than the distance of NGC 922 (43 Mpc). NGC 922 has a stellar mass of $M_* = 2.8 \times 10^{10}$ M$_\odot$ (P10) and a total dynamical mass of $M_{tot} = (9.1 \pm 0.8) \times 10^{10}$ M$_\odot$; hence, its gas to stellar mass fraction is $\log(M_{HI}/M_*) = -0.41 \pm 0.10$ and its gas to total mass fraction is $\log(M_{HI}/M_{tot}) = 0.11$.

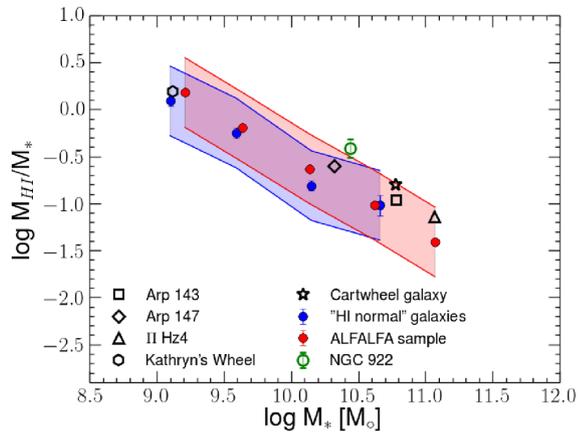

**Figure 7.** The averaged H I mass fraction of the 'H I-normal' galaxies (blue circles) presented in Cortese et al. (2011), of the H I stacked ALFALFA sample (red) presented in Brown et al. (2015) and of NGC 922 (green circle) as a function of stellar mass. The open square, diamond, triangle, hexagon and star are the gas fractions of the Arp 143, Arp 147, Π H z4, Kathryn's wheel and the Cartwheel galaxy, respectively (Marcum et al. 1992; Horellou et al. 1995; Higdon 1996; Romano et al. 2008; Parker et al. 2015). The blue and red shaded regions delimit the $y$-axis scatter in the 'H I-normal' galaxies and the ALFALFA sample, respectively.

Fig. 7 shows the average gas fraction scaling relationship as a function of stellar mass for the 'H I-normal' galaxies (blue), the H I stacked ALFALFA sample (red) and for NGC 922 (green circle). The open square, diamond, triangle, hexagon and star are the gas fractions for other collisional ring galaxies, namely the Arp 143, Arp 147, Π H z4, Kathryn's wheel and the Cartwheel galaxy, respectively (Marcum, Appleton & Higdon 1992; Horellou et al. 1995; Higdon 1996; Romano et al. 2008; Parker et al. 2015). The blue and red shaded areas are the average scatter from the mean values along the $y$-axis for the 'H I-normal' galaxies and the H I stacked ALFALFA sample, respectively. There is minor offset between the two samples, mainly owing to selection effects, that is, to differences in volume and the population sampled. However, the scatters in the averaged stellar mass to H I gas fractions are consistent. It is evident that

NGC 922 possesses a high amount of neutral hydrogen compared with galaxies of similar stellar mass (M$_\odot \sim 10^{10.4}$). The average H I gas fraction of these galaxies is $\log(M_{HI}/M_*) = -0.89$. However, it should be noted that the uncertainties in the gas fraction of NGC 922 are within the upper bounds of the scatter in these scaling relationships. In NGC 922, more than 80 per cent of the H I mass resides within its optical disc. The H I gas fraction of NGC 922, considering only the H I mass within its optical disc, is $\log(M_{HI}/M_*) = -0.50$, which is again higher than the average H I gas fraction of galaxies that have similar stellar masses. Furthermore, we find that other collisional ring galaxies have a higher H I gas fraction compared with the average H I gas fraction of galaxies with similar stellar masses. Hence, collisional ring galaxies have a higher H I gas fraction compared with the average gas fraction of 'H I-normal' galaxies at fixed stellar mass, but lie within the scatter of these scaling relationships.

Fig. 8 shows the H I mass-to-light ratio (based on the $R$-band luminosity) and SFR-to-H I mass maps of NGC 922. The SFR is estimated based on the H$\alpha$ luminosity of NGC 922 using the calibration of Meurer et al. (2009). We convolved all the optical images to match the beam size and the beam shape of our H I map. Hence, we note that the $M_{HI}/L_R$ and SFR/$M_{HI}$ ratios should be inferred within the region marked by the solid black ellipse, which represents the boundary of the optical radius (the star-forming C-shape morphology of NGC 922). Because colour gradients are common in spiral galaxies, gradients in the $M_{HI}/L_R$ ratio may also be present. Although the H I column density of NGC 922 is highest in the inner regions (see Fig. 3), the $M_{HI}/L_R$ ratio in the nucleus of this galaxy is the lowest (see the left panel of Fig. 8). We attribute this to the high $R$-band luminosity in the inner regions, in which the stellar densities are the highest compared with other parts of the galaxy. The $M_{HI}/L_R$ ratio is also low in the south-eastern corner of NGC 922, where there is a region with a high SFR apparent in the H$\alpha$ image (Fig. 1). The low $M_{HI}/L_R$ ratio in the south-eastern region suggests that this region is relatively more efficient in forming stars, which could also be deduced from the SFR/$M_{HI}$ map, as the SFR-to-H I mass ratio is high in this part of the galaxy. This interpretation is consistent with the finding that the south-eastern part of NGC 922 consists of four young star-forming complexes, with stellar ages of $\lesssim 7$ Myr and stellar masses of $(2 - 5) \times 10^6$ M$_\odot$ (P10). The mass of these four complexes combined comprises $\sim 60$ per cent of the total mass of the new star-forming complexes in this galaxy (with age $\lesssim 7$ Myr), which confirms the relatively high star-formation efficiency in this part of NGC 922.

The $M_{HI}/L_R$ ratio peaks in the northern part of NGC 922; however, this region is part of the northern H I tail, which lies beyond the optical disc of NGC 922 seen in the $R$-band image (see Fig. 3). This is more explicit in the SFR/$M_{HI}$ map, where the SFR-to-H I mass ratio is lowest in the northern and the western parts of the disc. In general, the SFR-to-H I mass ratios are highest in the inner, southern and eastern parts of the disc, which agrees well with the H$\alpha$ emission regions and the young cluster distribution found by P10. The majority of the young stellar clusters and the prominent H$\alpha$ emission regions (star-forming knots) are on the eastern and southern sides of the ring, while the western and northern sides (lowest SFR-to-H I mass ratio) host very faint H$\alpha$ emission regions. The total SFR-to-H I mass ratio (log SFR/$M_{HI}$) of NGC 922 is $-9.1 \pm 0.4$ yr$^{-1}$, which is high relative to the average value for the SINGG sample (log SFR/$M_{HI} = -9.6 \pm 0.3$ yr$^{-1}$), but still within the upper bounds of the scatter in this sample (Wong et al. 2016). In addition, the sSFR (sSFR = log SFR/$M_*$) of NGC 922 is $-9.5 \pm 0.3$ yr$^{-1}$, which places this galaxy above the average sSFR of galaxies with a similar stellar mass, but within the $1\sigma$ scatter (see e.g. Brinchmann et al. 2004; Abramson et al. 2014). Similarly, other ring galaxies occupy the same parameter space in the stellar mass $M_*$ versus sSFR diagram. For instance, the sSFR of Arp 147 is $-9.08$ yr$^{-1}$, which is again above the average sSFR of galaxies with the same stellar mass. The average SFR surface density of NGC 922 is $\log \Sigma_{SFR} = -2.30$ M$_\odot$ yr$^{-1}$ kpc$^{-2}$, and its average H I surface density is $\log \Sigma_{HI} = 0.91$ M$_\odot$ pc$^{-2}$, which places NGC 922 within the scatter of star-forming galaxies in the $\Sigma_{SFR}$–$\Sigma_{HI}$ parameter space (Bigiel et al. 2008; Schruba et al. 2011; Lemonias et al. 2014; Roychowdhury et al. 2015; Yim & van der Hulst 2016).

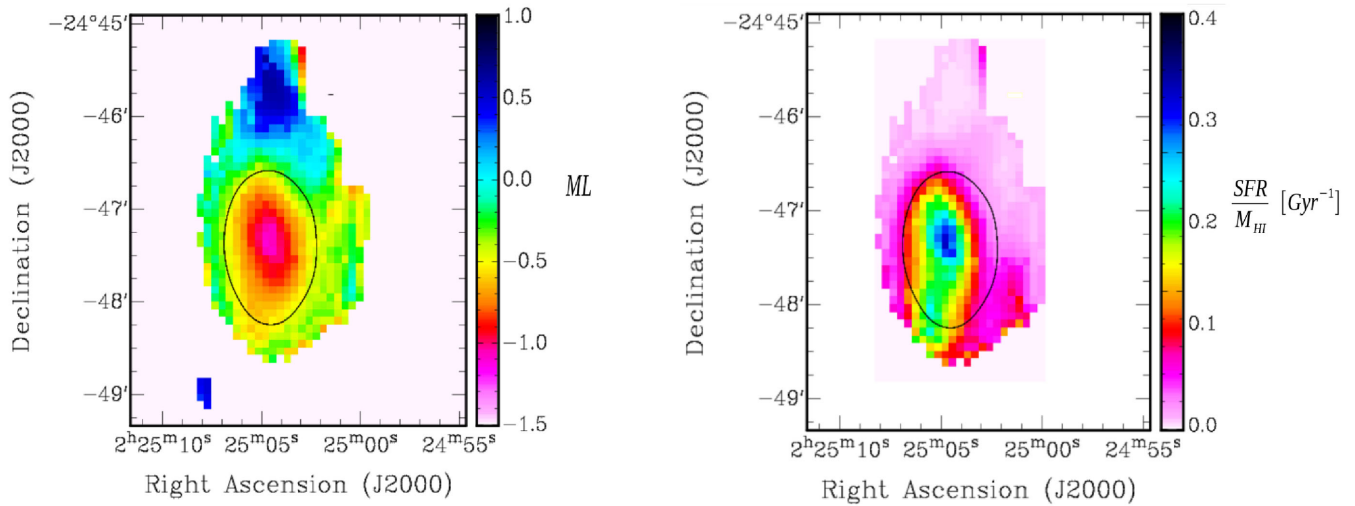

**Figure 8.** Left: the H I mass-to-light ratio ($ML = \log_{10}[M_{HI}/L_R]\,[M_\odot/L_\odot]$) image of NGC 922 in the $R$ band. Right: the SFR-to-H I mass ratio image estimated based on the H$\alpha$ luminosity of NGC 922 using the calibration of Meurer et al. (2009). The ellipse marks the star-forming C-shape morphology of NGC 922 after convolving the optical images to the beam size of our H I map.

At infrared wavelengths, NGC 922 has a high total infrared luminosity ($L_{TIR}$) compared with local starburst galaxies. Hanish et al. (2010) reported the total infrared luminosities of 13 local starburst galaxies and of 10 of their ordinary star-forming counterparts using the infrared bands of the Multiband Imaging Photometer of the *Spitzer* Space Telescope (MIPS; Werner et al. 2004; Rieke et al. 2004). The $L_{TIR}$ of NGC 922 ($\log L_{TIR} = 43.68 \pm 0.01$ erg s$^{-1}$) is higher than the values for both the starburst and non-starburst galaxies: the mean total infrared luminosity of starburst galaxies ($\log L_{TIR}$) is $42.2 \pm 1.1$ erg s$^{-1}$ and that for non-starburst galaxies is $41.7 \pm 0.9$ erg s$^{-1}$. In order to compare between different star formation indicators, we used the 24-μm MIPS band to trace the star formation in NGC 922. Using the calibration steps in Calzetti et al. (2007), the infrared-derived SFR (SFR$_{24\,\mu m}$) of NGC 922 is $8.5 \pm 0.6$ M$_\odot$ yr$^{-1}$, which is consistent with the FUV and the H$\alpha$ values reported in W06. We also compared the MIR–FIR flux ratios of the MIPS bands to trace the dust temperature in NGC 922. The 70-μm to 160-μm flux ratio of NGC 922 is twice the average value for the local star-forming galaxy samples presented in Draine et al. (2007) and Willmer et al. (2009). This high flux ratio indicates higher starlight intensities and hence hotter dust temperatures (Draine et al. 2007).

The collision process in ring galaxies triggers violent physical processes and coherent starburst activity that temporarily drive these systems to have higher SFRs than other galaxies with the same stellar mass. Similar starburst galaxies, for example ultra-luminous infrared galaxies (Sanders & Mirabel 1996), are believed to make a moderate contribution to the global SFR density, especially at higher redshifts (Rodighiero et al. 2011; Lamastra et al. 2013). Because interactions are more frequent at higher redshifts, as the rate of galaxy mergers increases with redshift (Bluck et al. 2012; Man, Zirm & Toft 2016), collisional ring galaxies are probably more abundant at these redshifts (Lavery et al. 2004). Thus, at higher redshifts collisional ring galaxies might contribute more to the cosmic SFR density than they do at zero redshift.

## 5 SUMMARY AND CONCLUSION

In this paper, we have presented our new H I observations of NGC 922. The high impact parameter and the speed of the intruder disrupted the H I morphology and induced warps in the gaseous disc of this galaxy. The gas warps are evident in the velocity field of NGC 922, and are more apparent in the approaching and the western side of the disc. Even though there is a plume of old stars emerging from this galaxy towards the intruder, we find no H I tail that is coincidental with the stellar plume. Instead, we find an H I tail that points in a different direction (north) with respect to the direction of the intruder S2 (north-west direction). This is inconsistent with the simple collisional scenario proposed for NGC 922 and is suggestive of additional tidal interactions that occurred before or after the drop-through of the intruder S2. Hence, our new H I observations provide greater constraints for future simulations and modelling of NGC 922.

We compared NGC 922 with unperturbed/isolated H I-rich systems in the local Universe to highlight its high gas fraction. The H I gas fraction in this galaxy is $\log(M_{HI}/M_*) = -0.41 \pm 0.10$, while the averaged H I gas fraction for galaxies with a similar stellar mass in our comparison sample is $\log(M_{HI}/M_*) = -0.9$. Furthermore, the high sSFR of NGC 922 places it above the average sSFR of galaxies with a similar stellar mass. Our results are consistent with NGC 922 having a SFR that is higher than the average SFR of galaxies with the same stellar mass. As we expect interactions to be more common at higher redshifts, collisional ring galaxies may account for a greater fraction of burst-dominated star-forming galaxies than they do in the local Universe.

## ACKNOWLEDGEMENTS

We thank the anonymous referee for their comments, which significantly improved the presentation of this manuscript. AE wishes to acknowledge the scholarship he received from the International Centre of Radio Astronomy (ICRAR). AE is also very grateful to T. Brown, T. Westmeier, B. Catinella, L. Cortese and D. Obreschkow for discussing some of the analysis done in this paper. This research has made use of the NASA/IPAC Extragalactic Database (NED), which is operated by the Jet Propulsion Laboratory, California Institute of Technology, under contract with the National Aeronautics and Space Administration. The Australia Telescope Compact Array is part of the Australia Telescope National Facility, which is funded by the Australian Government for operation as a National Facility managed by CSIRO. This paper includes archived data obtained through the Australia Telescope Online Archive (http://atoa.atnf.csiro.au).

## REFERENCES

Abramson L. E., Kelson D. D., Dressler A., Poggianti B., Gladders M. D., Oemler A., Jr, Vulcani B., 2014, ApJ, 785, L36
Allen R. J., Ekers R. D., Terlouw J. P., 1985, in di Gesu V., Scarsi L., Crane P., Friedman J. H., Levialdi S., eds, Data Analysis in Astronomy. Springer, Boston, MA, p. 271
Appleton P. N., Struck-Marcell C., 1996, Fund. Cosmic Phys., 16, 111
Barnes D. G. et al., 2001, MNRAS, 322, 486
Begeman K. G., 1987, PhD thesis, Kapteyn Institute
Begeman K. G., 1989, A&A, 223, 47
Bigiel F., Leroy A., Walter F., Brinks E., de Blok W. J. G., Madore B., Thornley M. D., 2008, AJ, 136, 2846
Bluck A. F. L., Conselice C. J., Buitrago F., Grützbauch R., Hoyos C., Mortlock A., Bauer A. E., 2012, ApJ, 747, 34
Braun R., Thilker D. A., 2004, A&A, 417, 421
Brinchmann J., Charlot S., White S. D. M., Tremonti C., Kauffmann G., Heckman T., Brinkmann J., 2004, MNRAS, 351, 1151
Brown T., Catinella B., Cortese L., Kilborn V., Haynes M. P., Giovanelli R., 2015, MNRAS, 452, 2479
Calzetti D. et al., 2007, ApJ, 666, 870
Chung A., van Gorkom J. H., Kenney J. D. P., Crowl H., Vollmer B., 2009, AJ, 138, 1741
Conn B. C., Fogarty L. M. R., Smith R., Candlish G. N., 2016, ApJ, 819, 165
Cornwell T. J., Uson J. M., Haddad N., 1992, A&A, 258, 583
Cortese L., Catinella B., Boissier S., Boselli A., Heinis S., 2011, MNRAS, 415, 1797
Draine B. T. et al., 2007, ApJ, 663, 866
Fiacconi D., Mapelli M., Ripamonti E., Colpi M., 2012, MNRAS, 425, 2255
Fogarty L. et al., 2011, MNRAS, 417, 835
Foster C. et al., 2014, MNRAS, 442, 3544
Gerber R. A., Lamb S. A., Balsara D. S., 1996, MNRAS, 278, 345
Ghosh K. K., Mapelli M., 2008, MNRAS, 386, L38
Giovanelli R. et al., 2005, AJ, 130, 2598
Hanish D. J., Oey M. S., Rigby J. R., de Mello D. F., Lee J. C., 2010, ApJ, 725, 2029
Haynes M. P., Giovanelli R., 1984, AJ, 89, 758
Higdon J. L., 1996, ApJ, 467, 241

Horellou C., Casoli F., Combes F., Dupraz C., 1995, A&A, 298, 743
Komatsu E. et al., 2011, ApJS, 192, 18
Koribalski B. S. et al., 2004, AJ, 128, 16
Lamastra A., Menci N., Fiore F., Santini P., 2013, A&A, 552, A44
Lavery R. J., Remijan A., Charmandaris V., Hayes R. D., Ring A. A., 2004, ApJ, 612, 679
Lemonias J. J., Schiminovich D., Catinella B., Heckman T. M., Moran S. M., 2014, ApJ, 790, 27
Lynds R., Toomre A., 1976, ApJ, 209, 382
Madore B. F., Nelson E., Petrillo K., 2009, ApJS, 181, 572
Man A. W. S., Zirm A. W., Toft S., 2016, ApJ, 830, 89
Mapelli M., Mayer L., 2012, MNRAS, 420, 1158
Marcum P. M., Appleton P. N., Higdon J. L., 1992, ApJ, 399, 57
Mayya Y. D., Bizyaev D., Romano R., Garcia-Barreto J. A., Vorobyov E. I., 2005, ApJ, 620, L35
Meurer G. R. et al., 2006, ApJS, 165, 307
Meurer G. R. et al., 2009, ApJ, 695, 765
Meyer M. J. et al., 2004, MNRAS, 350, 1195
Michel-Dansac L. et al., 2010, ApJ, 717, L143
Oh S.-H., Staveley-Smith L., Spekkens K., Kamphuis P., Koribalski B. S., 2018, MNRAS, 473, 3256
Parker Q. A., Zijlstra A. A., Stupar M., Cluver M., Frew D. J., Bendo G., Bojičić I., 2015, MNRAS, 452, 3759
Pellerin A., Meurer G. R., Bekki K., Elmegreen D. M., Wong O. I., Knezek P. M., 2010, AJ, 139, 1369 (P10)
Rieke G. H. et al., 2004, ApJS, 154, 25
Rodighiero G. et al., 2011, ApJ, 739, L40
Rogstad D. H., Lockhart I. A., Wright M. C. H., 1974, ApJ, 193, 309
Romano R., Mayya Y. D., Vorobyov E. I., 2008, AJ, 136, 1259
Roychowdhury S., Huang M.-L., Kauffmann G., Wang J., Chengalur J. N., 2015, MNRAS, 449, 3700
Sanders D. B., Mirabel I. F., 1996, ARA&A, 34, 749
Sault R. J., Teuben P. J., Wright M. C. H., 1995, in Shaw R. A., Payne H. E., Hayes J. J. E., eds, Astronomical Society of the Pacific Conference Series Vol. 77, Astronomical Data Analysis Software and Systems IV. Astron. Soc. Pac., San Francisco, p. 433
Schruba A. et al., 2011, AJ, 142, 37
Struck-Marcell C., Lotan P., 1990, ApJ, 358, 99
Theys J. C., Spiegel E. A., 1977, ApJ, 212, 616
van der Hulst J. M., Terlouw J. P., Begeman K. G., Zwitser W., Roelfsema P. R., 1992, in Worrall D. M., Biemesderfer C., Barnes J., eds, Astronomical Society of the Pacific Conference Series Vol. 25, Astronomical Data Analysis Software and Systems I. Astron. Soc. Pac., San Francisco, p. 131
Werner M. W. et al., 2004, ApJS, 154, 1
Willmer C. N. A., Rieke G. H., Le Floc'h E., Hinz J. L., Engelbracht C. W., Marcillac D., Gordon K. D., 2009, AJ, 138, 146
Wilson W. E. et al., 2011, MNRAS, 416, 832
Wong O. I. et al., 2006, MNRAS, 370, 1607 (W06)
Wong O. I., Meurer G. R., Zheng Z., Heckman T. M., Thilker D. A., Zwaan M. A., 2016, MNRAS, 460, 1106
Yim K., van der Hulst J. M., 2016, MNRAS, 463, 2092
Yun M. S., Ho P. T. P., Lo K. Y., 1994, Nature, 372, 530
Zwaan M. A. et al., 2004, MNRAS, 350, 1210

This paper has been typeset from a TEX/LATEX file prepared by the author.